\documentstyle[prd,aps]{revtex} 

\input psfig \pssilent
\begin{document}
\title{Discrete quantum gravity: applications to cosmology}

\author{Rodolfo Gambini$^1$, and Jorge Pullin$^2$}
\address{1. Instituto de F\'{\i}sica, Facultad de Ciencias, 
Universidad
de la Rep\'ublica\\ Igu\'a esq. Mataojo, CP 11400 Montevideo, Uruguay}
\address{2. Department of Physics and Astronomy, 
Louisiana State University,\\ 202 Nicholson Hall, Baton Rouge,
LA 70803-4001}

\date{December 7th. 2002}
\maketitle
\begin{abstract}
We consider the application of the consistent lattice quantum gravity
approach we introduced recently to the situation of a Friedmann
cosmology and also to  Bianchi  cosmological models.  This allows us
to work out in detail the computations involved in  the determination of
the Lagrange multipliers that impose consistency, and the implications
of this determination.  It also allows us to study the removal of the
Big Bang singularity.  Different discretizations can be achieved
depending on the version of the classical theory chosen as a starting
point and their relationships studied. We analyze in some detail how the
continuum limit arises in various models. In particular we notice how
remnants of the symmetries of the continuum theory are embodied in 
constants of the motion of the consistent discrete theory. The 
unconstrained nature of the discrete theory allows the consistent
introduction of a relational time in quantum cosmology, free from
the usual conceptual problems.
\end{abstract}

\section{Introduction}

The use of lattice regularizations in quantum gravity has always been
problematic due to the fact that the introduction of the lattice
breaks diffeomorphism invariance (see Loll \cite{loll} for a review).
The lattice theory therefore has a very different symmetry group than
the continuum theory it approximates. We have recently introduced
\cite{us} a procedure to treat in a consistent way the resulting
discrete theories, including their symmetries, both at a classical and
quantum mechanical level. As is common in discrete theories, the
resulting theory fixes Lagrange multipliers that in the continuum
theory are free \cite{TDLee}. This is in principle puzzling and it is
worthwhile examining in detail in situations where one can easily
control all aspects. In this paper we study these issues in the
context of  cosmological models: the Friedmann and the Bianchi 
models. These models have several appealing properties: they have
enough degrees of freedom as to have a set of non-trivial observables and 
make the deparameterization and physical 
description of the evolution possible, which is a feature one
would like to analyze in the full theory. The Bianchi II model is
classically soluble through a canonical transformation that maps it to
a free particle. One can discretize the model before or after the
canonical transformation. The discrete theories are not the same and
there is the question as of their similarities and
differences. Moreover, generically, other canonical transformations
(like the one that introduces the Ashtekar variables) in the continuum
will also lead to different theories upon discretization. We will
see that the discrete theories avoid the singularity and have 
symmetries that translate themselves in constants of the motion
that are remnants of the observables of the continuum theory. We
will see that the discrete theory allows naturally to introduce
a relational notion of time free of the usual problems of such
approach.

We have recently introduced \cite{us} a general technique for treating 
the theories that arise when one discretizes a continuum field
theory. We have shown that the technique works for Yang--Mills and
BF theories and implemented it for the gravitational case. Here
we summarize the basic features of the approach. Since lattice
field theories have a finite number of degrees of freedom, and
since in this paper we are looking at cosmologies only, it 
suffices to summarize for illustration purposes the technique
when applied to a constrained mechanical system.

Consider a constrained mechanical system. We assume we replace
in the action all time derivatives with discrete first order
finite differences. The time integral in the action is replaced
by a sum $S=\sum_{n=1}^N L(q_n,q_{n+1})$ with,
\begin{equation}
L(n,n+1) = p_n (q_{n+1}-q_n) -\epsilon H(q_n,p_n)-\lambda_{nB} \phi^B(q_n,p_n)
\end{equation}
where we assume we have a Hamiltonian $H$ and $M$ constraints $B=1\ldots M$
and $\epsilon$ is the time interval. Our formulation works for a system with
an arbitrary (finite) number of phase-space degrees of freedom. To simplify
the notation we are omitting the indices labeling the degrees of freedom
and writing formulae as if the system only had one configuration degree of
freedom.

The centerpiece of our technique is to realize that in a theories where
time (and space in the case of field theories) is discretized, it makes no
sense to work with a Hamiltonian since the latter is the generator of 
infinitesimal time evolution and one cannot have infinitesimal evolution 
if time is discrete. The time evolution should be described via a canonical
transformation that implements the discrete time evolution between instants
$n$ and $n+1$. A type 1 canonical transformation that accomplishes this
task has as generating function minus the Lagrangian, viewed as a function
of $q_n$ and $q_{n+1}$. From it, one 
obtains the momenta at instant $n+1$
\begin{eqnarray}
P^q_{n+1} &=& {\partial L(n,n+1) \over \partial q_{n+1}} =p_n\label{37}\\
P^p_{n+1}&=& {\partial L(n,n+1) \over \partial p_{n+1}}= 0\label{38}\\
P^{\lambda_B}_{n+1}&=& {\partial L(n,n+1) \over \partial \lambda_{(n+1)B}}= 0,
\end{eqnarray}
and similarly, the momenta at instant $n$ are given by,
\begin{eqnarray}
P^q_n &=& -{\partial L(n,n+1) 
\over \partial q_{n}} =p_n+\epsilon {\partial H(q_n,p_n) \over \partial q_n} + \lambda_{nB}
{\partial \phi^B(q_n,p_n) \over \partial q_n}\label{40}\\
P^p_n &=& -{\partial L(n,n+1) 
\over \partial p_{n}} = -(q_{n+1}-q_n) + \epsilon {\partial H(q_n,p_n) \over \partial p_n} 
+ \lambda_{nB}
{\partial \phi^B(q_n,p_n) \over \partial p_n}\label{41}\\
P^{\lambda_B}_n &=& \phi^B(q_n,p_n).\label{42}
\end{eqnarray}
We can now combine these equations to give,
\begin{eqnarray}
p_n-p_{n-1} &=& -\epsilon {\partial H(q_n,p_n) \over \partial q_n} - \lambda_{nB}
{\partial \phi^B(q_n,p_n) \over \partial q_n}\label{43}\\
q_{n+1}-q_n &=&
\epsilon {\partial H(q_n,p_n) \over \partial p_n} 
+ \lambda_{nB} {\partial \phi^B(q_n,p_n) \over \partial p_n}\label{44}\\
\phi^B(q_n,p_n) &=&0.\label{45}
\end{eqnarray}
Superficially, these equations appear entirely equivalent to the continuum ones.
However, they hide the fact that in order for the constraints to be preserved,
the Lagrange multipliers get fixed. Another way to see it, is that $
P^q_{n+1} =p_n$ and therefore it is immediate that the Poisson bracket of the
constraints evaluated at $n$ and at $n+1$ is non-vanishing. It is illuminating
to rewrite these equations in terms of the canonically conjugate pair,
\begin{eqnarray}
P^q_{n+1}-P^q_{n} &=& -\epsilon {\partial H(q_n,P^q_{n+1}) \over \partial q_n} - \lambda_{nB}
{\partial \phi^B(q_n,P^q_{n+1}) \over \partial q_n}\label{11}\\
q_{n+1}-q_n &=&
\epsilon {\partial H(q_n,P^q_{n+1}) \over \partial P^q_{n+1}} 
+ \lambda_{nB} {\partial \phi^B(q_n,P^q_{n+1}) \over \partial P^q_{n+1}}\label{12}\\
\phi^B(q_n,P^q_{n+1}) &=&0.\label{13}
\end{eqnarray}

We therefore will eliminate the ``constraints'' (\ref{13}) by solving
them for the Lagrange multipliers. One can proceed in various ways,
either determining the Lagrange multipliers at time $n$ as functions
of the variables at either $n,n-1$ or $n+1$.  We will choose to
determine the Lagrange multipliers at instant $n$ as  functions of
$P^q_n$ and $q_n$ in this paper. For different problems it can be more
convenient to make one choice over the other. In all cases solving for
the Lagrange multipliers implies solving algebraic equations, but
depending on the choices made the resulting equations can be quite
non-linear and guaranteeing that they will have real solutions can be
problematic. For the examples of this paper we have found that the
choice we make is the most convenient one.

We solve (\ref{11}) for $P^q_{n+1}$ and substitute it in the ``constraint'' (\ref{13})
and obtain 
\begin{equation}
\phi^B(q_{n},P^q_{n},\lambda_{nB})=0, \label{system}
\end{equation}
and this constitutes a system of equations. Generically, these will 
determine 
\begin{equation}
\lambda_{nB} =\lambda_{nB}(q_{n},P^q_{n}, v^\alpha) \label{lambdadet}
\end{equation}
where
the $v^\alpha$ are a set of free parameters that may arise if the system of
equations is undetermined. The eventual presence of these parameters will signify that
the resulting theory still has a genuine gauge freedom represented by freely
specifiable Lagrange multipliers. 

The final set of evolution equations for the system is therefore given
by (\ref{11},\ref{12}) where the Lagrange multipliers are substituted
using (\ref{lambdadet}).

This construction raises several questions about how to implement it
in concrete gravitational examples. We would like to probe these
questions in the context of cosmological models. The main points we
want to probe are the following:

{\em (i) Solubility of the multiplier equations:} Solving the 
constraints by choosing the Lagrange multipliers produces a theory that
is constraint free. This is analogous to what happens when one gauge
fixes a theory. It is well known that gauge fixing is not a cure for the
problems of general relativity since gauge fixings usually become 
problematic. In the same sense, it could happen that the algebraic
equations that determine the Lagrange multipliers (the lapse and the shift 
in the case of general relativity) in our approach develop problems in
their solutions (for instance, negative lapses, or complex solutions).
We will show that in these simple examples there do not appear to 
be difficulties even in situations where a traditional gauge fixing in
the continuum theory is not possible.

{\em (ii) Performing meaningful comparisons:} When comparing a
discrete theory with a continuum theory, one needs to choose the
quantities that are to be compared. In particular, the continuum
theory has ``observables'' (``perennials''), that is, quantities that
have vanishing Poisson brackets with the constraints.  The discrete
theory is constraint-free. We will see that one has exact conserved
quantities in the discrete theory that are consequences of remnant
symmetries encoded in the initial data that the discrete theory
inherits from the continuum theory. The conserved quantities will
turn out to be closely related to the observables of the continuum
theory.

{\em (iii) The continuum limit}: in continuum constrained theories
with first class constraints, the Lagrange multipliers are free functions.
Yet in our discrete construction, the Lagrange multipliers are determined
by the initial conditions. If one wishes to take a naive continuum limit,
the discrete equations that determine the Lagrange multipliers must become
ill-defined. To extract meaningful information about the continuum theories,
one needs to proceed differently. We have already mentioned that the discrete
theories are constraint-free, since one solves the constraints for the 
Lagrange multipliers. This in particular means that they have more 
degrees of freedom than the continuum theory they are trying to approximate. 
The continuum limit might be achieved by a careful fine tuning of the 
extra degrees of freedom, or might occur in certain asymptotic regions of
the dynamics as an attractor for a large set of initial values of the extra
degrees of freedom. We will exhibit examples of both kinds of behaviors.

{\em (iv) Singularities:} Discrete theories in principle have the
possibility of evolving through a Big Bang singularity, since generically
the singularity will not lie on a point on the lattice. However, we will
see that if one uses canonical variables such that the singularity is at
a boundary of the range of a variable, then the discrete theories do 
develop singularities, although they can be avoided in certain cases.

{\em (v) Problem of time:} We will discuss how to obtain 
evolution in the discrete by using a relational approach in terms of 
the observables of these theories. This suggests that the problem of
time can be solved in these theories.

{\em (vi) Discretization ambiguities}:
An important element is to note that the Lagrange multipliers get
determined by this construction only if the constraint 
is both a function of $q$ and
$p$. If the constraint is only a function of $q$ or of $p$ then the
constraints are automatically preserved in evolution without fixing
the Lagrange multipliers. This raises a conceptual question.  For
certain theories in the continuum, one can make a canonical
transformation to a new set of variables such that the constraints
depend only on $q$ or on $p$. The resulting discrete theories will
therefore be very different in nature, but will have the same
continuum limit. From the point of view of using discrete theories to
quantize gravity, we believe this ambiguity should receive the same
treatment that quantization ambiguities (choice of canonical
variables, factor orderings, etc.)  get: they are decided
experimentally. Generically there will be different discrete theories
upon which to base the quantization and some will be better at
approximating nature than others in given regimes. Many of them may 
allow to recover the continuum limit, however they may have 
different discrete properties when one is far from the semiclassical regime.

The organization of this paper is as follows: in the next section we
discuss several Friedmann models that exhibit various behaviors upon
discretization. In section III we discuss the quantization of the
models and the various possibilities concerning how one approaches the
``problem of time''. In section IV we discuss other models, including
the Bianchi ones and show that they exhibit similar kinds of behaviors
as the ones we encountered in the Friedmann models. We end with a
discussion of the importance of the results.

\section{Friedmann models}
\subsection{Continuum theory}
We will consider a Friedmann cosmological model, written in terms of
Ashtekar's variables \cite{kodama}. The fundamental canonical pair is
$(E,A)$ where $E$ is the only remnant of the triad after the
minisuperspace reduction and $A$ is its canonically conjugate
variable. We will consider the presence of a cosmological constant and
of a scalar field. We will assume the scalar field has a very large
mass so we can neglect its kinetic term in the Hamiltonian constraint,
for the sake of computational simplicity. The Lagrangian for the model
is,

\begin{equation}
L= E\dot{A} + \pi \dot{\phi}- N E^2 (-A^2+(\Lambda +m^2 \phi^2)|E|)\label{lag}
\end{equation}
where $\Lambda$ is the cosmological constant, $m$ is the mass of the
scalar field $\phi$, $\pi$ is its canonically conjugate momentum and
$N$ is the lapse with density weight minus one. The appearance of
$|E|$ in the Lagrangian is due to the fact that the term cubic in $E$
is supposed to represent the spatial volume and therefore should be
positive definite. In terms of the ordinary lapse $\alpha$ we have
$\alpha = N |E|^{3/2}$. The equations of motion and constraint are
\begin{eqnarray}
\dot{A} &=& N (\Lambda+m^2\phi^2) {\rm sgn}(E) E^2\\
\dot{E} &=&2 N E^2 A\\
\dot{\phi}&=&0\\
\dot{\pi} &=&-2 N |E|^3 m^2 \phi\\
A^2&=&(\Lambda+m^2\phi^2) |E|\label{constr}
\end{eqnarray}

It immediately follows from the large mass approximation that $\phi={\rm constant}$.
To solve for the rest of the variables, we need to distinguish four cases,
depending on the signs of $E$ and $A$. Let us call $\epsilon={\rm sgn}(E)$ and
$\chi={\rm sgn}(A)$. Then the solution (with the choice of lapse $\alpha=1$) is,
\begin{eqnarray}
A&=&\chi \exp\left(\chi\epsilon t \sqrt{\Lambda+m^2\phi^2} \right)\\
E&=&\epsilon {\exp\left(2\chi\epsilon t \sqrt{\Lambda+m^2\phi^2} \right) 
\over {\Lambda+m^2\phi^2} }
\end{eqnarray}

There are four possibilities according to the signs  $\epsilon,\chi$.
If $\epsilon=\chi=1$ or $\epsilon=\chi=-1$ we have a universe that expands.
If both have different signs, the universe contracts. This just reflects that
the Lagrangian is invariant if one changes the sign of either $A$ or $E$
and the sign of time. It is also invariant if one changes simultaneously
the sign of both $A$ and $E$.

Let us turn to the observables of the theory (quantities that have vanishing
Poisson brackets with the constraint (\ref{constr}) and therefore are constants
of the motion). The theory has four phase space degrees of freedom with one
constraint, therefore there should be two independent observables. Immediately
one can construct an observable $O_1=\phi$, since the latter is conserved due
to the large mass approximation. To construct the second observable we write
the equation for the trajectory,
\begin{equation}
{d\pi \over d A}= {-2 E m^2 \phi \over \Lambda+m^2 \phi^2} =
-{2 A^2 m^2 \phi \over \left(\Lambda+m^2 \phi^2\right)^2} {\rm sgn} E
\end{equation}
where in the latter identity we have used the constraint. Integrating, we
get the observable,
\begin{equation}
O_2 = \pi + {2 \over 3} {m^2 \phi \over \left(\Lambda+m^2
\phi^2\right)^2} A^3 {\rm sgn} E \label{obse1}
\end{equation}
and using the constraint again we can rewrite it,
\begin{equation}
O_2 = \pi + {2 \over 3} {m^2 \phi \over \Lambda +m^2 \phi^2} A E.
\end{equation}

Although the last two expressions are equivalent, we will see that upon
discretization only one of them becomes an exact observable of the discrete
theory.
\subsection{Discrete theory}

We consider the evolution parameter to be a discrete variable. Then the
Lagrangian becomes
\begin{equation}
L(n,n+1)=E_n (A_{n+1}-A_n)+ \pi_n (\phi_{n+1}-\phi_n) -N_n E_n^2 (-A^2_n+
(\Lambda+m^2 \phi^2_n) |E_n|)
\end{equation}

The discrete time evolution is generated by a canonical transformation of
type 1 whose generating function is given by $-L$, viewed as a function
of the configuration variables at instants $n$ and $n+1$. The canonical
transformation defines the canonically conjugate momenta to all variables.
The transformation is such that the symplectic structure so defined is 
preserved under evolution. The configuration variables will be
$(A_n,E_n,\pi_n,\phi_n,N_n)$ with canonical momenta 
$(P^A_n,P^E_n,P^\phi_n,P^\pi_n,P^N_n)$, defined by,
\begin{eqnarray}
P^E_{n+1} &=& {\partial L(n,n+1) \over \partial E_{n+1}} =0,\\
P^E_{n} &=& -{\partial L(n,n+1) \over \partial E_{n}} =
-(A_{n+1}-A_n)+N_n E^2_n(\Lambda+m^2 \phi^2_n){\rm sgn} E_n,\\
P^A_{n+1}&=& {\partial L(n,n+1) \over \partial A_{n+1}}= E_n,\\
P^A_{n}&=& -{\partial L(n,n+1) \over \partial A_{n}}= 
E_n-2 N_n A_n E_n^2,
\\
P^\phi_{n+1}&=& {\partial L(n,n+1) \over \partial \phi_{n+1}}= \pi_n,\\
P^\phi_{n}&=& -{\partial L(n,n+1) \over \partial \phi_{n}}= \pi_n
+2 N_n E_n^2 m^2 \phi_n |E_n|,
\\
P^\pi_{n+1}&=& {\partial L(n,n+1) \over \partial \pi_{n+1}}= 0,\\
P^\pi_{n}&=& -{\partial L(n,n+1) \over \partial \pi_{n}}= -\phi_{n+1}+\phi_n,\\
P^N_{n+1}&=& {\partial L(n,n+1) \over \partial N_{n+1}}= 0,\\
P^N_{n}&=& -{\partial L(n,n+1) \over \partial N_{n}}= 
E_n^2 (-A^2_n+ (\Lambda+m^2 \phi^2_n) |E_n|).
\end{eqnarray}
These equations can be recast into a more familiar looking fashion,
by  combining the information at
levels $n$ and $n+1$,
\begin{eqnarray}
A_{n+1}-A_n&=&N_n (P^A_{n+1})^2_n(\Lambda+m^2 \phi^2_n){\rm sgn} P^A_{n+1},
\label{anm1}\\
P^A_{n+1}-P^A_n &=&2 N_n A_n  (P^A_{n+1})^2\label{pnm1}\\
\phi_{n+1}-\phi_n&=&0\label{phinm1}\\
P^\phi_{n+1}-P^\phi_n &=&-2 N_n (P^A_{n+1})^2 m^2 \phi_n |P^A_{n+1}|,\\
0&=&-A^2_n+ (\Lambda+m^2 \phi^2_n) |P^A_{n+1}|,\label{cons2}
\end{eqnarray}
and the phase space is now spanned by $A_n,P^A_n,\phi_n,P^\phi_n$. 

Enforcing the constraint (\ref{cons2}) leads to the determination of the
Lagrange multipliers. There are two ways to proceed. We can use the
evolution equation (\ref{pnm1}) to eliminate $P^A_{n+1}$ in the constraint.
The latter will therefore be an equation that determines $N_n$ as a 
function of $P^A_n,A_n,\phi_n$. The alternative is to use (\ref{anm1},\ref{phinm1})
to eliminate $A_{n}$ and $\phi_n$. This would yield $N_n$ as a function of
$P^A_{n+1},\phi_{n+1},A_{n+1}$. We will here follow the first approach since
it is the more natural one to track evolution forward in the parameter $n$.

From (\ref{pnm1}) we determine,
\begin{equation}
P^A_{n+1}={1 +\xi \sqrt{1-8P^A_n A_n N_n}\over 4 A_n N_n},
\end{equation}
where $\xi=\pm 1$ and we will see the final solution is independent of $\xi$.
Substituting in (\ref{cons2}) and solving for the lapse we get,
\begin{equation}
N_n = {\left[-P^A_n \left(\Lambda+m^2 \phi_n^2\right)+A_n^2{\rm sgn}P^A_n\right]
\left(\Lambda+m^2\phi_n^2\right)\over 2 A_n^5}.\label{lapse}
\end{equation}

Let us summarize how the evolution scheme presented actually
operates. Let us assume that some initial data $A^{(0)},P^A_{(0)}$, 
satisfying the constraints of the continuum theory, are
to be evolved. The recipe will consist of assigning $A_0=A^{(0)}$ and
$P^A_1=P^A_{(0)}$. Notice that this will automatically satisfy (\ref{cons2}).
In order to the scheme to be complete we need to
specify $P^A_0$. This is a free parameter. Once it is specified, then
the evolution equations will determine all the variables of the
problem, including the lapse. Notice that if one chooses $P^A_0$ such
that, together with the value of $A_0$ they satisfy the constraint,
then the right hand side of the equation for the lapse (\ref{lapse})
would vanish and no evolution takes place. It is clear that one can
choose $P^A_0$ in such a way as to make the evolution step as small as
desired.

The equation for the lapse (\ref{lapse}) implies that the lapse is a
real number for any real initial data. But it does not immediately
imply that the lapse is positive. However, it can be shown that the
sign of the lapse, once it is determined  by the initial configuration,
does not change under evolution. The proof is tedious since one has to
separately consider the various possible signs of $\epsilon$ and $\chi$.
Let us exhibit the proof for the case $\epsilon=\chi=1$. Then $A_n>0$ 
and $P_n>0$. To simplify
the notation let us set $\Theta=\Lambda+m^2\phi^2$, which is a positive
quantity. The equation determining the lapse becomes,
\begin{equation}
N_n={\left[-P_n^A\Theta+A^2_n\right]\Theta\over 2 A_n^5}.
\end{equation}
Let us assume we start with a positive lapse, i.e., $A_n^2/\Theta>P_n$.
The equations of motion then imply $A_{n+1}>A_n$ and $P^A_{n+1}>P^A_n$.
If we now compute the lapse in the next instant of time we get,
\begin{equation}
N_{n+1}={\left[-P_{n+1}^A\Theta+A_{n+1}^2\right]\Theta \over 2 A_{n+1}^5}=
{\left[-A_n^2+A_{n+1}^2\right]\over 2 A_{n+1}^5}>0
\end{equation}
where in the last step we used the constraint. Notice that if one had
chosen initial values such that $N_n<0$, then $A_{n+1}<A_n$ and $P_{n+1}<P_n$ 
and the equation for is $N_{n+1}<0$. In effect, one is following the same
trajectory backwards in time, i.e. the universe contracts instead of 
expanding. A similar analysis holds for the other choices of $\epsilon$
and $\chi$.

This is an important result. In spite of the simplicity of the model, there was
no a priori reason to expect that the construction would yield a real lapse.
Or that upon evolution the equation determining the lapse could not become singular
or imply a change in the sign of the lapse, therefore not allowing a complete
coverage of the classical orbits in the discrete theory.

We would like to address the issue of how to compare the discrete
theory with the continuum one. Notice that this is a priori a delicate
proposition. The continuum theory has a constraint and therefore a
gauge symmetry. The discrete theory is constraint-free. A direct
comparison of a given variable in the continuum theory with its discrete
counterpart is therefore not adequate, since one could be comparing
a solution with its counterpart in not necessarily the same gauge.
This suggests that a more meaningful comparison could be attempted if
one considered the observables of the continuum theory, since the latter
are gauge independent. Even this is a priori problematic. In the discrete
theory, since there are no constraints, any quantity is an observable.
One could consider quantities in the discrete theories that arise as 
discretizations of the observables of the continuum theory. But due to
discretization ambiguities, there are many discrete counterparts to each
observable in the continuum. All of them are ``observables'' of the 
discrete theory since there are no constraints. Is there any that is
preferred? We will show that indeed one can find discretizations of the
observables of the continuum theory that are exact constants of the motion of
the discrete theory.

In order to exhibit this, we start by rewriting evolution equations
that result from determining the lapse as discussed above. We make
them explicit for $\epsilon=1$,
\begin{eqnarray}
P^A_{n+1}&=&A_n^2 \Theta^{-1}\label{ecupa}\\
A_{n+1}&=&{3A_n^2-P^A_n\Theta\over 2A_n}\label{ecua}\\
\phi_{n+1}&=&\phi_n\\
P^\phi_{n+1}&=&P^\phi_n -\left(A_n^3-P^A_n\Theta A_n\right)m^2\phi_n\Theta^{-2}
\end{eqnarray}
where we recall that $\Theta=\Lambda+m^2\phi_n^2$. It should be noted that
these equations preserve the symplectic structure, that is, the variables
$(P^q_{n+1},q_{n+1})$ have the same canonical Poisson brackets as $(P^q_n,q_n)$.
The resulting dynamical system has {\em four phase space} degrees of freedom.
This is in contrast to the system in the continuum, that had four phase space
degrees of freedom with one first class constraint, resulting in only {\em two
phase space degrees of freedom}. This appears as surprising, since one seems
to be attempting to approximate a theory with two degrees of freedom with
a discrete theory that has four degrees of freedom. To understand this better,
let us consider the issue of the observables of the continuum theory and their
discrete counterpart. We will see that the discrete theory appears to have 
hidden symmetries that may help explain why it approximates correctly a
theory with in principle a different number of degrees of freedom.

Let us recall the continuum expression of the non-trivial observable we found,
\begin{equation}
O = \pi + {2 \over 3} {m^2 \phi \over \Lambda +m^2 \phi^2} A E.
\end{equation}
If we consider one possible discretization of this expression,
\begin{equation}
O_n=P^\phi_n+{2 \over 3} {m^2 \phi_n \over \Theta_n} A_n P^A_n,
\end{equation}
we can check that the discrete equations of motion immediately imply
$O_{n+1}=O_n$.  That is, this expression is an exact constant of the
motion of the discrete theory.  This is a rather remarkable result. It
is not at all true that a generic discretization of an observable of
the continuum theory will yield a constant of the motion of the
discrete theory. For instance, if we had started from (\ref{obse1})
and discretized it, one would not get a constant of the motion. A
further ambiguity arises in the fact that we have made the choice for
all elements of the right hand-side to be at the level $n$ when
discretizing.

It is also remarkable that the discrete theory would have constants of
the motion, and more remarkable that such constants of the motion can
be obtained by discretization of the observables of the continuum theory.
We do not know if this is a generic feature of discretized theories or
if it just is something that occurs in some  examples. 

In general the discrete theory, having more degrees of freedom than
the continuum theory, will have more constants of the motion than 
observables in the continuum theory. In this example, the discrete
theory has four degrees of freedom. One can find four constants of the
motion. One of them we already discussed. The other one is $\phi$. 
The two other constants of the motion can in principle be worked
out. One of them is a measure of how well the discrete theory 
approximates the continuum theory and is only a function of the 
canonical variables (it does not depend explicitly on $n$). The
constant of the motion is associated with the canonical transformation
that performs the evolution in $n$. It is analogous to the Hamiltonian
of the discrete theory. The expression can be worked out as a power
series involving the discrete expression of the constraint of the 
continuum theory. This constant of motion therefore vanishes in
the continuum limit. The other constant of the motion also vanishes
in the continuum limit.

That is, we have two of the constants of the motion that reduce to the
observables of the continuum theory in the continuum limit and two others
that vanish in such limit. The discrete theory therefore clearly has a
remnant of the symmetries of the continuum theory. The canonical 
transformations associated with the constants of the motion which 
have non-vanishing continuum limit map dynamical trajectories to other
trajectories that can be viewed as different choices of lapse in the
discrete theory. This is the discrete analog of the reparameterization
invariance of the continuum theory. As we will see in the next 
section the lapse in the discrete theory is determined up to two
constants. The choice of these two constants is the remnant of the
reparameterization invariance of the continuum theory that is present
in the discrete theory.

\subsection{Comparing the discrete and the continuum theories}
\label{compar}

We would like to compare the predictions of the discrete theory with those of the 
continuum theory. Since we are comparing two theories that are different in nature,
one has to provide a mapping between them. We start by identifying the variables
$t=\epsilon n$ in the continuum and discrete theories respectively where $\epsilon$
is a real positive quantity. The discrete theory provides a unique solution given
initial data, including a lapse function. We therefore should cast the continuum
theory with a lapse that is close to the one generated by the discrete theory,
otherwise the comparison would not be meaningful. In order to do this, we
first derive a recurrence relation for the variable $A_n$. We obtain it by 
combining eqs. (\ref{ecupa},\ref{ecua}),
\begin{equation}
A_{n+1} = {3A_n^2-A_{n-1}^2\over 2A_n}.\label{recur}
\end{equation}
We now write an equation in the continuum such that its discretization would
yield this recursion relation,
\begin{equation}
2A\ddot{A}+\dot{A}^2=0,
\end{equation}
where we approximate $\ddot{A}=(A_{n+1}+A_{n-1}-2A_n)/\epsilon^2$ and
$\dot{A}=(A_{n+1}-A_n)/\epsilon$. The solution of the differential equation is,
$A=C_A (t+k)^{2/3}$ for $t+k>0$ where $C_A$ and $k$ are constants that will
depend on the initial values specified to the recursion relation $A_0$ and 
$A_1$. This corresponds to one of the four
branches we considered before. If we now consider the constraint, this would
determine $E$ as a function of the parameter $t$. Substituting $E$ and $A$ 
in the continuum evolution equations one obtains an algebraic equation for  
the densitized lapse $N$. The solution is $N=C_N/(t+k)^3$, the ordinary
lapse is $\alpha=C_\alpha/(t+k)$ where $C_N$ and $C_\alpha$ are again
constants that are determined by the initial values of the recursion
relation. We therefore see that the discrete theory has a remnant of the
reparameterization symmetry of the continuum theory:
in the continuum limit it can reproduce a two parameter family of 
functional forms of the lapse.

\begin{figure}
\centerline{\psfig{file=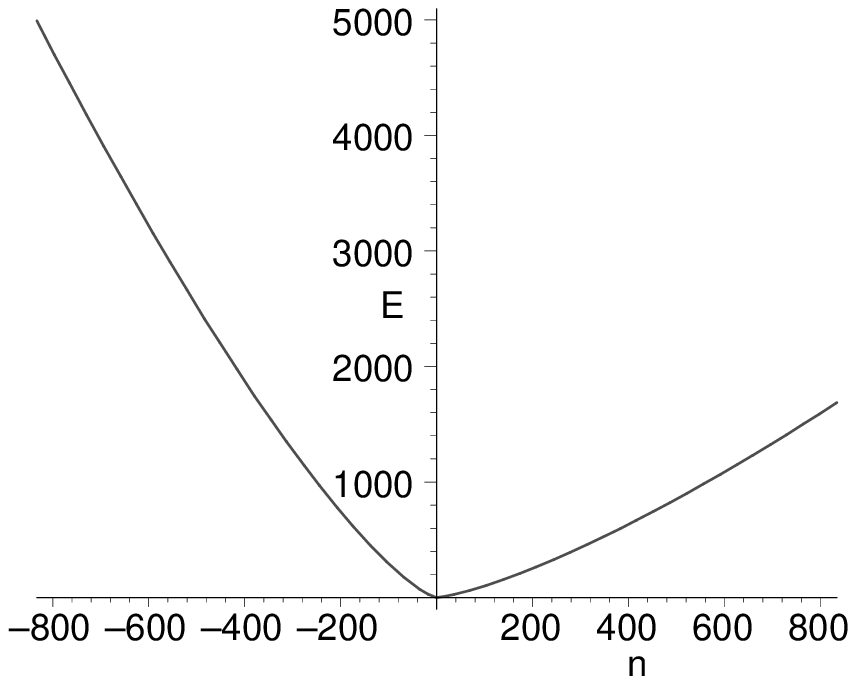,height=5cm},\psfig{file=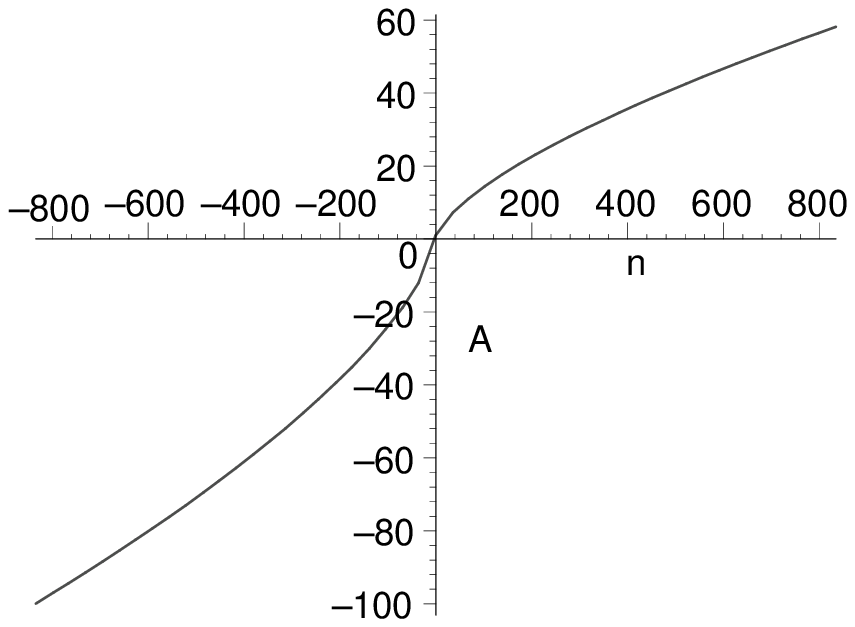,height=5cm}}
\caption{The discrete evolution of the triad $E$ and the connection $A$ as a
function of the discrete evolution parameter $n$. We have chosen initial 
conditions that produce a positive branch of $A$ for $n>0$ and a negative
branch for $n<0$. For the triad we chose both branches positive.}
\label{eadis}
\end{figure}

Figure (\ref{eadis}) shows the discrete evolution of $E$ and $A$. The branch
chosen is such that one has a universe that first contracts and then expands.
We have chosen the parameter in such a way that the point where one would 
expect the singularity occurs at $n=0$. 

\begin{figure}
\centerline{\psfig{file=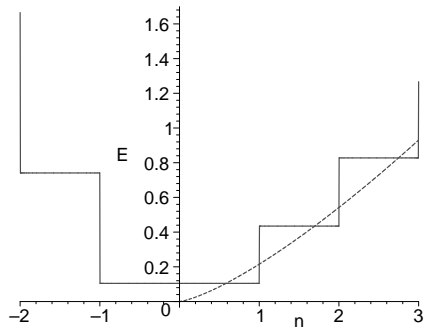,height=5cm}}
\caption{The approach to the singularity in the discrete and continuum 
solutions. The discrete theory has a small but non-vanishing triad at $n=0$
and the singularity is therefore avoided.}
\label{edism}
\end{figure}

Although the graphs suggest that the triad goes to zero at $n=0$ and
therefore one has a singularity, this is not the case. Figure (\ref{edism})
shows in more detail the approach to $n=0$. It also includes superposed on
it the continuum solution. As we see, the continuum solution indeed goes
to zero at $t=0$. The discrete solution takes a small but non-zero value.
One could choose initial data such that the discrete solution becomes 
singular, but it would correspond to a set of measure zero of initial
data. Generically, the singularity is avoided in the discrete theory.

One can also notice that while going through the singularity the
universe re-expands, but the behavior at both sides of the singularity
is different. In this sense the discrete cosmology may implement the
proposal of Smolin \cite{Smolin} that different physics takes place
when one goes through a singularity.

\begin{figure}
\centerline{\psfig{file=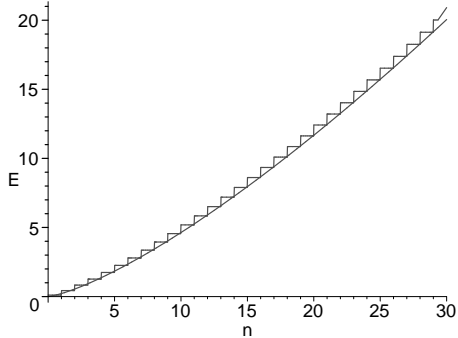,height=5cm}}
\caption{Agreement between the continuum and discrete solutions for the triad.}
\label{edismed}
\end{figure}

In figure (\ref{edismed}) we show the level of agreement between the discrete
and continuum solutions for the triad. If we take a larger scale the two
curves are indistinguishable.

\begin{figure}
\centerline{\psfig{file=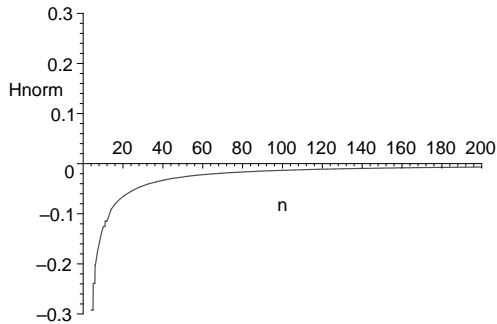,height=5cm}}
\caption{A normalized measure of the Hamiltonian constraint of the
continuum theory (see text for details), evaluated in the discrete
theory. We see that the constraint is better satisfied at later times,
indicating that the discrete theory is achieving a continuum limit.}
\label{ham}
\end{figure}

A better idea of the agreement between the continuum and the discrete theory
can be obtained by evaluating the Hamiltonian constraint of the continuum
theory in the discrete theory. The constraint reads $A^2-E\Theta=0$, if 
we discretize it as $A^2_n-P^A_n \Theta_n$ this in general is not zero in
the discrete theory, what vanishes is equation (\ref{cons2}), which is 
different. Figure (\ref{ham}) shows a measure of how well the constraint of the
continuum theory is satisfied. In order to normalize the result in a meaningful
way (since the quantity is supposed to be zero) we plot $H_{norm}=1-A^2_n/P^A_n\Theta$.

The presence of the two constants of motion in the discrete theory
implies two relations between the dynamical variables of the
theory. If one combines these relations with the constraint of the
continuum theory, one will obtain further relations among the
dynamical variables that will hold only approximately, since the
constraint of the continuum theory is not exactly satisfied in the
discrete theory. However, due to the results shown in figure
\ref{ham}, these relations will approximate well the relations that
appear in the continuum theory as a consequence of the observables
present in such theories, and their combination with the constraint of
the continuum theory. It is to be noted that the latter are
relationships among the variables of the continuum theory that are
parameterization independent.  The agreement between the continuum and
discrete theory at the level of these relations indicates that both
are agreeing at a gauge invariant level.

\begin{figure}
\centerline{\psfig{file=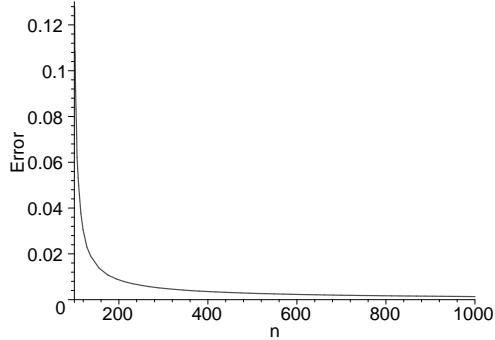,height=5cm}}
\caption{The difference between the value of the variable $P^\phi_n$ calculated
directly from the evolution equations and by inferring it from the equation
that defines the observable in the continuum theory. As one enters the region
where the discrete theory describes the continuum well this difference goes
to zero. We plot the difference divided by the value of $P^\phi$ since the 
latter grows and therefore the difference also increases in value, but 
decreases in relative value.}
\label{figobs}
\end{figure}

As an example of this, in figure \ref{figobs} we display the difference
between the value of the variable $P^\phi_n$ computed directly from 
the evolution of the discrete theory and its value computed using
the continuum relation (\ref{obse1}) translated to the discrete theory,
\begin{equation}
P^\phi_n=O_2-
{2 \over 3} {m^2 \phi_n \over \left(\Lambda+m^2\phi_n^2\right)^2} A_n^3.
\end{equation}
Since $P^\phi_n$ grows as $n^2$ in this example, we show in the figure
the difference between the $P^\phi_n$ calculated both ways divided by
the norm of the exact value of $P^\phi_n$.

\section{Quantization}

One of the main attractive features of the consistent discrete lattice approach
is that the resulting discrete theory is free of constraints. This implies that
many of the deep conceptual problems of quantum gravity appear to be eliminated
from the outset. As we will see, this is not necessarily the case. We will
see that the quantization of the model is more subtle that what one would 
initially attempt and that there are several possibilities for quantization,
depending on the attitude one adopts with respect to the ``problem of time''.

\subsection{Naive quantization}

Given that the theory is unconstrained, and that time evolution is
represented by a canonical transformation, it appears that one can
readily quantize the system. One starts by picking a polarization, for
instance wavefunctions $\Psi[A,\phi]$.  It is natural to operate in
the Heisenberg picture. In this picture operators are labeled by the
``time level'' $n$.  Operators $\hat{A}_0$ and $\hat{\phi}_0$ act
multiplicatively and $\hat{P^A}_0=i{\partial /\partial A}$ and
$\hat{P^\phi}_0=i {\partial/\partial \phi}$.  Wavefunctions are square
integrable functions $A$ and $\phi$ and the fundamental operators are
self-adjoint on this space. Operators at other instances of ``time''
are obtained via a unitary transformation. This is guaranteed by the
fact that in the classical theory the discrete time evolution was
implemented via a canonical transformation. The usual problem of the
dynamics of a canonical quantization, i.e.  implementing the
Hamiltonian as an operator in a certain factor ordering translates
here into finding a unitary operator that implements at an operatorial
level the classical discrete equations of motion for the system.

To construct the unitary operator we start by writing the following 
matrix elements,
\begin{equation}
<p,\phi_1,n| U |A,\phi_2,n>=<p,\phi_1,n+1| A,\phi_2,n>,
\end{equation}
where $<p,\phi,n|$ and $|A,\phi,n>$ are bases of states at instant $n$
labeled by the Eigenvalues of $\hat{\phi}_n$ and $\hat{P^A}_n$ and
$|A,\phi,n>$ respectively, which we denote by $p$ and $A$.

We now consider the matrix elements of the 
operatorial version of equation (\ref{ecupa}),
\begin{equation}
<p,\phi_1,n+1|\hat{|P^A_{n+1}|}-\hat{A}_n^2 \hat{\Theta}_n^{-1}|A,\phi_2,n>=
\left(|P^A|-A^2 \Theta^{-1}\right) <p,\phi_1,n+1|A,\phi_2,n>=0.
\end{equation}
This implies that 
\begin{equation}
<p,\phi_1,n+1|A,\phi_2,n>=f(A,\phi_1,\phi_2) \delta\left(|p|-A^2 \Theta^{-1}\right)
\end{equation}
with $f(A,\phi_1,\phi_2)$ an arbitrary normalization factor. This factor can
be determined by studying in a similar fashion the other evolution equations.
 The final result
for the unitary evolution operator is,
\begin{eqnarray}
<p,\phi_1,n+1||A,\phi_2,n>&=&\sqrt{2 |A|\over \Theta}
\delta\left(p-{A^2\over \Theta} \right)\delta(\phi_1-\phi_2)
\exp\left(i {A^3\over \Theta}\right)H(A)\\ &+& \sqrt{2 |A|\over
\Theta} \delta\left(p+{A^2\over \Theta} \right)\delta(\phi_1-\phi_2)
\exp\left(-i {A^3\over \Theta}\right)H(-A)\nonumber
\end{eqnarray}
where $H(A)$ is the Heaviside function $H(A)=1 \forall A>0$, zero otherwise.

One can Fourier transform to obtain the matrix elements in the $|A,\phi,n>$ 
basis,
\begin{eqnarray}
<A_1,\phi_1,n+1|A_2,\phi_2,n>&=&
\sqrt{2 |A_2|\over \Theta} \exp\left(-i {A^2_2(A_1-A_2)\over \Theta}
\right)\delta(\phi_1-\phi_2) H(A_2)\\
&+&
\sqrt{2 |A_2|\over \Theta} \exp\left(i {A^2_2(A_1-A_2)\over \Theta}
\right)\delta(\phi_1-\phi_2) H(-A_2).
\end{eqnarray}

Dirac \cite{Dirac} had already noted in 1933 that the unitary operator
that implements a canonical transformation is given by $\exp(-i G)$
where $G$ is the generating function of the canonical
transformation. In our case the generating function (after eliminating
the Lagrange multipliers is indeed given by $G(A_{n},A_{n+1})=
A^2_{n}(A_{n+1}-A_{n})/\Theta$.  There is an overall difference with
Dirac's result since he chooses a specific factor ordering that does
not coincide with the one we chose. It is interesting that this 
construction is what led Dirac to the notion of path integral.

With this unitary transformation one can answer any question about
evolution in the Heisenberg picture for the model in question. One
could also choose to work in the Schr\"odinger picture, evolving the
states. Notice that the wavefunctions admit a genuine probabilistic
interpretation. This is in distinct contrast with the usual 
``naive Schr\"odinger interpretation'' of quantum cosmology which
attempted to ascribe probabilistic value to the square of a solution
of the Wheeler--DeWitt equation (see \cite{Kuchar} for a detailed
discussion of the problems associated with the naive interpretation).

An interesting aspect of this quantization is that for any
square-integrable wavefunction and for any value of the parameter $n$,
the expectation value of $(P^A_n)^2$, and therefore that of $E^2$ is
non-vanishing, and so are the metric and the volume of the
slice. Therefore quantum mechanically one never sees a singularity.
This was expected, classically the singularity only occurred for a set
of measure zero of initial conditions in the discrete theory and
therefore quantum mechanically, since we are superposing many initial
conditions, the singularity has zero probability. We therefore have a
similar prediction to the one Bojowald \cite{Bojobojo} encountered in
his approach to quantum cosmology in the loop representation, but the
details of how the singularity is avoided are quite different.  For
instance, in Bojowald's approach there is an instant in the evolution
in which the volume goes through zero and nevertheless the inverse
scale factor is finite. The metric is not a well defined operator at
all. Here the volume never goes through zero, the inverse scale factor
and the metric are always finite.

Although the attempted quantization is complete, its interpretation as
a quantum theory of cosmology, is problematic. This has to do with the
fact that the ``evolution'' variable $n$ does not have any intrinsic
meaning.  Classically there is no combination of the variables of the
problem that would tell us what the value of $n$ is. There is no well
defined notion of ``future'' and ``past'' since a generic quantum
state will be a superposition of expanding and contracting universes
(recall that in this approach the initial data determine if the
universe expands or contracts). Therefore the quantum theory, although
complete, is lacking in predictive power.

\subsection{``Gauge fixed'' quantization}

The formulation of the theory we considered for quantization in the last
section is such that the same orbit of the continuum theory is represented
by many orbits in the discrete theories, dependent on the initial data
specified, as we discussed in section \ref{compar}. One could therefore
attempt a quantization after eliminating such remnant symmetry. This 
would be tantamount to a ``gauge fixed'' quantization. In order to
do this we choose two constants $T^0,T^1$ and set $T^0=A_0$ and $T^1=A_1$,
which also determines $P^A_0=3(T^0)^2-2T^0T^1/\Theta$. We can from this
eliminate completely the variables $A_n$ and $P^A_n$ from the theory
and use the variable $T^n=A^n$ as a time variable.

The theory now has as variables $\phi_n,P^\phi_n$. The Lagrangian of
the theory (\ref{lag}) becomes
\begin{equation}
L(n,n+1)=\pi_n (\phi_{n+1}-\phi_n) +{(T^{n-1})^2\over \Theta_n}
(T^{n+1}-T^n)
\end{equation}
where the last term is obtained by evolving the equations of motion and 
obtaining $A$ and $E$ as function of the initial data and $\phi$. The
equations of motion for the theory are generated by a canonical 
transformation generated by $-L$ as usual. Eliminating some of the 
variables, we get as end result,
\begin{eqnarray}
\phi_{n+1}&=&\phi_n,\\
P^\phi_{n+1}-P^\phi_n &=& {\left[(T^{n-1})^2 T^n-(T^n)^3\right]m^2 
\phi_n \over \Theta^2}.
\end{eqnarray}

To quantize the theory we consider a Hilbert space of square
integrable functions $\Psi[\phi,T^n]$, the actions of the operators
$\hat{\phi}$ and its canonically conjugate are the obvious ones. One
can introduce a unitary operator that implements the canonical
transformation. In this theory, the variable $n$ determines, through
$T^n$, the value of a time variable that has the physical meaning of
the connection $A_n$. Notice that we now really have a two-parameter
family of quantum theories. The predictions of these theories are in
general not equivalent, but they all share the same continuum limit.
In particular, some of the theories (for instance, choose initial
data such that $P^0=0$) generically include a singularity,
in clear contrast to what we observed in the naive quantization.

Ignoring the peculiarities introduced by the nature of the discrete
theory, what we have attempted here is to solve the problem of time in
cosmology by choosing one of the dynamical variables as time at the
classical level and therefore keep it classical when performing the
quantization. Again, this is aided by the fact that we do not have
constraints.

\subsection{Relational time quantization}

The choice of a given variable as classical in order to define a
``time'' is motivated from the ordinary quantization of
non-relativistic mechanical systems, but it is highly unnatural in the
context of quantum gravity. In ordinary quantum mechanics, it is
assumed that ``t'' is measured by a classical clock external to the
system. Since when quantum gravity effects are important one cannot
expect to have classical clocks available, the most natural
possibility to introduce a concept of time is to treat all variables
quantum mechanically and use one of them as a clock as long as it
behaves semi-classically.  This is an old idea, for instance already
described in DeWitt's original article \cite{Dewitt} on canonical
quantum gravity. Page and Wootters \cite{page} described the idea in
detail in the context of ordinary quantum mechanical systems. The
application to quantum gravity of this idea usually runs into
problems, as clearly discussed by Kucha\v{r} \cite{Kuchar}.  In
ordinary canonical quantum gravity problems arise when one chooses the
variables. One needs variables that change during the evolution.
Therefore their corresponding quantum operators cannot commute with
the constraints. As a consequence, they are not well defined operators
on the space of states annihilated by the constraints. One could try
to work on the space of all kinematical states, where the operators
would be well defined.  But the solutions to the constraints on such a
space are distributional and therefore cannot be used to construct a
probabilistic interpretation, which is needed to define the
correlations of variables one wishes to introduce when considering one
variable as time.

We will see that in our approach one can indeed introduce a notion
of relational time without confronting the difficulties that appear
in ordinary canonical quantization. Since our discrete theory is 
constraint free, one can consider any variable of the theory as a 
time variable and study correlations with other variables. The only
requirement will be that the variable be a ``good clock'', in the 
sense that the variable exhibits a semi-classical behavior. There 
might be regimes in which no variable satisfies the requirements and
no notion of semi-classical time can therefore be introduced in such
regimes.

To define a time we therefore introduce the conditional probabilities,
``probability that a given variable have a certain value when
the variable chosen as time takes a given value''. 
For instance, taking $A$ as our time variable, let us work out first
the probability
that the scalar field conjugate momentum be in the range 
$\Delta P^\phi= [P^\phi_{(1)},P^\phi_{(2)}]$
and ``time'' is in the range $\Delta A=[A_{(1)}, A_{(2)}]$
(the need to work with ranges is because we are dealing with continuous
variables). We go back to the naive quantization and recall that the
wavefunction $\Psi[A,\phi,n]$ in the Schr\"odinger representation 
admits a probabilistic interpretation.
One can also define the amplitude $\Psi[A,P^\phi,n]$ by taking the
Fourier transform. Therefore the probability of simultaneous measurement
is, 
\begin{equation}
P_{\rm sim}(\Delta P^\phi,\Delta A) = \lim_{N\to \infty} {1 \over N}
\sum_{n=0}^N \int_{P^\phi_{(1)},A_{(1)}}^{P^\phi_{(2)},A_{(2)}}
\Psi^2[A,P^\phi,n] dP^\phi dA. \label{psim}
\end{equation}
We have summed over $n$ since there is no information about the ``level''
of the discrete theory at which the measurement is performed, since $n$
is just a parameter with no physical meaning. With the normalizations
chosen if the integral in $P^\phi$ and $A$ 
were in the range $(-\infty,\infty)$, $P_{\rm sim}$ would be equal to one.

To get the conditional probability $P_{\rm cond}(\Delta P^\phi|\Delta A)$,
that is, the probability that having observed $A$ in $\Delta A$ we also
observe $P^\phi$ in $\Delta P^\phi$, we use the standard probabilistic
identity 
\begin{equation}
P_{\rm sim}(\Delta P^\phi,\Delta A) =
P(\Delta A) P_{\rm cond}(\Delta P^\phi|\Delta A)
\end{equation}
where $P(\Delta A)$ is obtained from expression (\ref{psim})
taking the integral on $P^\phi$ from $(-\infty,\infty)$. We therefore
get 
\begin{equation}
P_{\rm cond}(\Delta P^\phi|\Delta A)=
{\lim_{N\to \infty} {1 \over N}
\sum_{n=0}^N \int_{P^\phi_{(1)},A_{(1)}}^{P^\phi_{(2)},A_{(2)}}
\Psi^2[A,P^\phi,n] dP^\phi dA \over
\lim_{N\to \infty} {1 \over N}
\sum_{n=0}^N \int_{-\infty,A_{(1)}}^{\infty,A_{(2)}}
\Psi^2[A,P^\phi,n] dP^\phi dA}.  
\end{equation}
Notice that all the integrals are well defined and the resulting 
quantity behaves as a probability in the sense that integrating 
from $(-\infty,\infty)$ in $P^\phi$ one gets unity.

Introducing probabilities is not enough to claim to have completed
a quantization. One needs to be able to specify what happens to the
state of the system as a measurement takes place. The most natural
reduction postulate is that,
\begin{equation}
|\Psi> = \lim_{N\to\infty}  \sum_0^N \Psi(A,P^\phi,n) |A,P^\phi,n>.
\end{equation}
This yields the status of the state after measuring $P^\phi$ in the 
instant of ``time'' $A$.

One should discuss in more detail the construction of the quantum
theory. In particular, a question that might be raised is if the 
relational approach pursued here reproduces ordinary quantum mechanics
in simpler systems. This is discussed in detail in a separate publication 
\cite{porto}, where we show how to recover ordinary quantum mechanics
with this approach, in appropriate regimes.

\section{Other models}

In this section we will discuss other cosmological models we have
analyzed. In the interest of brevity we will not work them out 
at the same level of detail as the one we presented in the previous
sections but we will comment about similarities and differences.

\subsection{Closed Friedmann model with $\Lambda<0$ and very massive scalar field}

In the previous sections we discussed the cosmology of a flat 
Friedmann model coupled to a very massive scalar field and with a positive
cosmological constant. Here we would like to consider a model with
spatial curvature $k=-1$, negative cosmological constant and coupled to a 
very massive scalar field. The classical continuum theory of these models
predicts the universe recollapses.

The continuum Lagrangian is
\begin{equation}
L=E\dot{Q}+\pi \dot{\phi}-N E^2\left(-Q^2+(\Lambda+m^2\phi^2)|E|+{1 \over 4}\right),
\end{equation}
where we have chosen to write it in terms of purely real variables
corresponding to the ``triad'' $E$ and its canonically conjugate
momentum $Q$. In terms of the Ashtekar connection $A=Q+i/2$ where $i$
the imaginary unit \cite{kodama}

The equations of motion are,
\begin{eqnarray}
\dot{Q} &=&N\Theta\, {\rm sgn}(E) E^2\\
\dot{E} &=&2 N E^2 Q\\
\dot{\phi} &=&0\\
\dot{\pi} &=& -2 N |E|^3 m^2 \phi\\
0&=&-Q^2+|E|\Theta +{1 \over 4} 
\end{eqnarray}
where as before $\Theta=(\Lambda+m^2\phi^2)$. The range of $\phi$ is
restricted to $\phi^2<-\Lambda/m^2$ so that $\Theta<0$, otherwise the
universe will not recollapse.  The last equation is the Hamiltonian
constraint and has as immediate consequence  that $|Q|<1/2$ and 
that $|E|<-1/(4\Lambda)$. If one fixes the lapse $\alpha=N |E|^{3/2}$ to
one, one gets the following classical solution,
\begin{eqnarray}
Q&=& {1 \over 2} \cos\left(\sqrt{-\Theta} t\right),\\
E&=& -{1 \over 4\Lambda} \sin^2\left(\sqrt{-\Theta} t\right),
\end{eqnarray}
which reproduces the well known recollapsing Friedmann universe.

The following observables have vanishing Poisson brackets with the 
Hamiltonian constraint,
\begin{eqnarray}
O_1&=&\phi\\
O_2&=& \pi +\left[{2\over 3} {Q^3 m^2 \phi \over \Theta^2} -
{Q m^2 \phi \over 2 \Theta^2}\right]{\rm sgn}(E).
\end{eqnarray}

Discretization of the theory is straightforward and works along the
same lines as in the $k=0$ case. The recurrence relation for the real
part of the connection $Q$ remains the same as (\ref{recur}). As in
the previous case, there exist exact constants of the motion of the
discrete theory, given by $\phi_n$ and
\begin{equation}
O_2= P^\phi_n+\left[{2\over 3} {Q_n P^A_n m^2 \phi_n \over
\Theta_n^2} - {Q_n m^2 \phi_n \over 3 \Theta^2_n}\right],
\end{equation}
for ${\rm sgn}(E)=1$. 

The behavior of the discrete model in its approximation to the 
continuum theory is different than in the $k=0$ case. In that
case the approach to the continuum was asymptotic for large 
values of the parameter $n$. In the $k=1$ model such a behavior
cannot be possible since the model starts recollapsing. The
discrete theory has a chance of better approximating the continuum
theory in some intermediate region between the points at which
the Big Bang and the Big Crunch would occur in the continuum 
theory. This is not generic, however. If one chooses initial
data that imply too long a time step in evolution, it might be
that one does not reach the continuum limit by the time one is
at the Big Crunch.

\begin{figure}
\centerline{\psfig{file=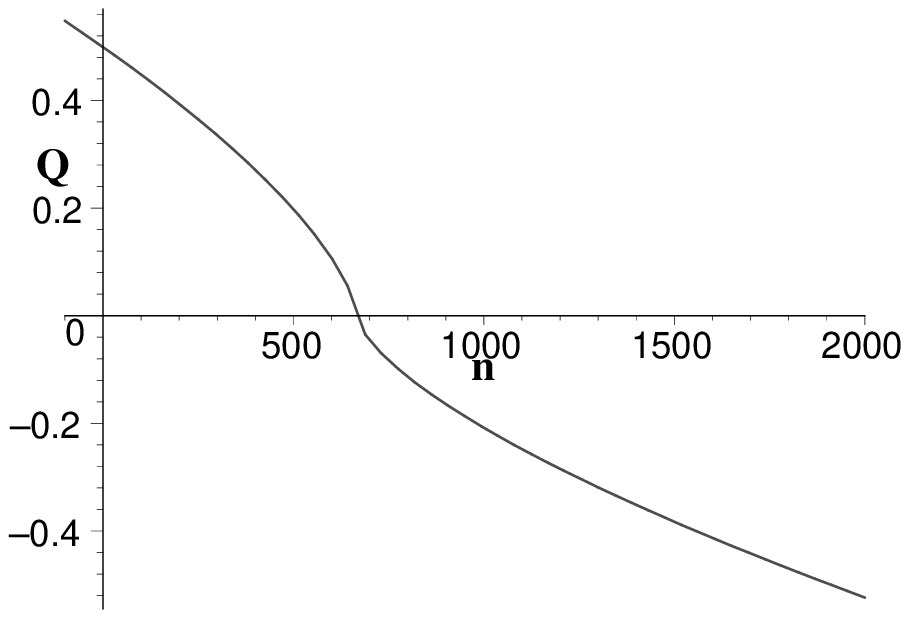,height=5cm},\psfig{file=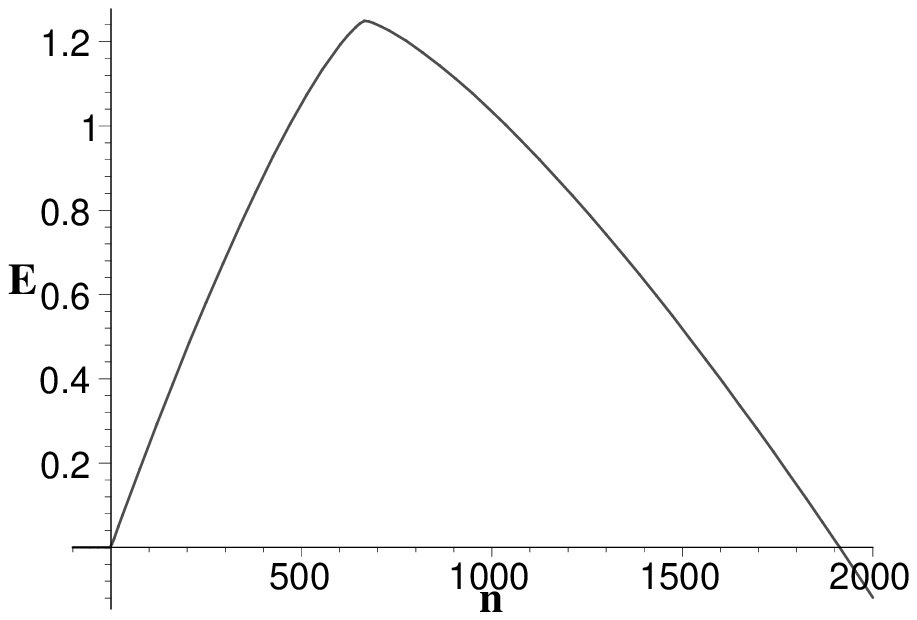,height=5cm}}
\caption{The evolution of the real part of the Ashtekar
connection $Q$ and the triad $E$ for the discrete closed
Friedmann model. The kink that is observed at the moment
of maximum expansion corresponds to the fact that at that point
the connection goes through zero and the discrete theory picks
up two different parameterizations of the continuum theory 
at both sides of $Q=0$. It is therefore a coordinate feature.
If one looks at the observables of the theory, they remain
constant throughout evolution and nothing special happens
at that point.}
\label{fig1km1}
\end{figure}

\begin{figure}
\centerline{\psfig{file=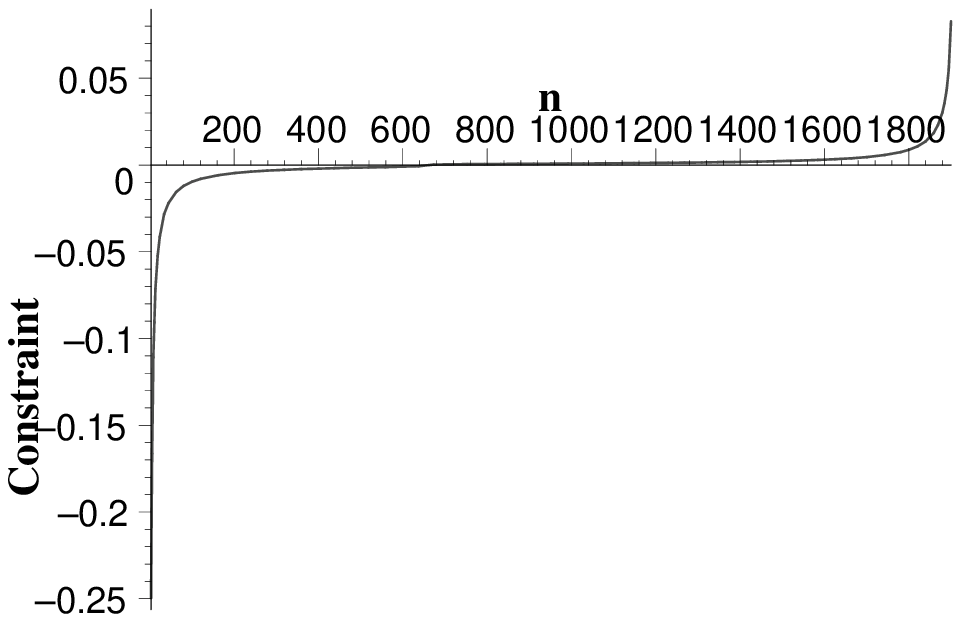,height=5cm}}
\caption{The continuum Hamiltonian constraint of the continuum theory
written in the dimensionless form $1+(-Q^2 +{1 \over 4})/(|E|\Theta)$
evaluated on the discrete theory for the closed Friedmann model. The
constraint is close to zero in the region of maximum expansion showing
that the model achieves best approximation of the continuum theory in
that region.}
\label{fig2km1}
\end{figure}

Figure \ref{fig1km1} shows the evolution for the real part of
the Ashtekar connection $Q$ and the triad $E$ for the discrete
closed Friedmann model. The initial data were chosen in such 
a way that the continuum limit is well approximated in the
intermediate maximum expansion region, as shown in figure 
\ref{fig2km1}

\subsection{Flat Friedmann model with $\Lambda=0$ and massless scalar field}

This model will teach us about situations in which the continuum limit is not
achieved as in the previous models. 

The continuum Lagrangian is
\begin{equation}
L=E\dot{A}+\pi \dot{\phi}-N \left(-E^2 A^2+\pi^2\right).
\end{equation}

The equations of motion are straightforward. 
The model has two independent observables,
\begin{eqnarray}
O_1&=& A \exp\left(\epsilon \phi\right),\label{o1cont}\\
O_2&=& E \exp\left(-\epsilon \phi\right),\label{o2cont}
\end{eqnarray}
and the obvious observable $\pi$ can be obtained from them as $\pi =\epsilon O_1 O_2$,
where $\epsilon =\pm 1$. Discretizing the action we get,
\begin{eqnarray}
L=E_n \left(A_{n+1}-A_n\right) + \pi_n \left(\phi_{n+1}-\phi_n\right) -N_n 
\left(-E^2_n A_n^2+\pi_n\right).
\end{eqnarray}

Working out the equations of motion, the Lagrange multiplier gets determined to,
(choosing $\epsilon=+1$),
\begin{equation}
N_n={1\over 2} {P^\phi_n-A_n P^A_n\over (P^\phi_n)^2}.
\end{equation}
The resulting equations of motion are,
\begin{eqnarray}
P^A_{n+1}&=&{P^\phi_n\over A_n},\label{85}\\
A_{n+1}&=& {A_n^2 P^A_n \over P^\phi_n},\label{86}\\
\phi_{n+1} &=& \phi_n+{P^\phi_n-A_n P^A_n\over P^\phi_n},\label{87}\\
P^\phi_{n+1} &=&P^\phi_n.
\end{eqnarray}

Combining (\ref{85}) and (\ref{86}) one gets a recursion relation,
\begin{equation}
{A_{n+1}\over A_n}={A_n \over A_{n-1}}=a.
\end{equation}

Given initial values for $A_0$ and $A_1$, the solution to the
recursion relation is immediate, stating that the ratio of two
successive connections is a constant $a$.  The solution is therefore
of a power law form: $A_n=A_0 a^n$, $P^A_n=P^A_0 a^{-n}$.

Combining (\ref{85}) and (\ref{87}) we get,
\begin{equation}
\phi_{n+1} =\phi_n + 1-a.
\end{equation}

To recover a continuum limit we therefore need to take the limit $a\to 1$.
Let us therefore consider $a=\exp(-\delta)$ with $\epsilon$ close to zero,
we get $\phi_{n+1}-\phi_n=\delta$ and $\phi_n=\phi_0 +n\delta$. If we 
obtain $\delta$ from this last expression and substitute it in the solution
for $A_n$ we get $A_n=\exp(\phi_n-\phi_0) A_0$. Therefore $A_n \exp(-\phi_n)$
is a constant of the motion which reproduces the observable of the continuum
theory (\ref{o1cont}). If one computes the normalized Hamiltonian constraint
of the continuum theory evaluated on the discrete theory one gets,
\begin{equation}
1 - {A_n P^A_n \over P^\phi_n} = \delta,\label{viola}
\end{equation}
and indeed one recovers the continuum theory for $\delta\to 0$.

Notice that the continuum limit in this model arises in a very different way
than in the previous model. There we had that for any initial data one 
asymptotically into the future reached a region that approximated well the
continuum theory. The continuum limit was therefore an ``attractor'' in the
dynamics of the theory. In this model one sees that the level of 
approximation of the continuum theory, as exhibited by the violation of the
Hamiltonian constraint (\ref{viola}) is constant throughout the evolution.
To achieve a certain level of accuracy one will have to choose the
initial data such that the value of $\delta$ is the desired one.

This behavior however, has negative implications at the time of quantization.
Since a quantum state is generically a superposition of many possible initial
conditions, there will be no way of achieving a classical limit, unless one
fine tunes the initial state carefully so only initial data that approximate
the classical theory are included in the superposition. In the previous 
examples the universe generically behaved in a quantum way near the Big Bang
and then became more classical into the future. In this model the universe
will behave in the same fashion throughout the complete evolution, independent
of time.

\subsection{The Bianchi cosmologies}

We will not present a full discussion of Bianchi cosmologies, but will
just mention some salient features in which some of the models resemble
the behavior of the Friedmann models discussed before, and one situation
in which they differ significantly.

The Bianchi cosmologies are spatially homogeneous space-times where
the group of homogeneities acts transitively. If the structure
constants of the algebra of symmetries are traceless, then one has the
``Bianchi class A'' models (Bianchi I, II, VIII, IX) \footnote{For simplicity
we omit the class $VI_0$ model, which is also class A ??}. These models
admit a straightforward Hamiltonian formulation as a reduction of the
full action of general relativity via the symmetry. It is well known
that the dynamics of these cosmological models classically is
equivalent to a particle bouncing inside a potential well. In the case
of the Bianchi IX and VIII the well is closed and the particle executes
an infinite series of bounces against the walls. Each individual bounce
can be approximated by a Bianchi type II solution and the motion of the
particle between bounces can be approximated by a Bianchi type I 
solution. So basically by studying the Bianchi I and II models we 
capture approximately all the features needed to study the more 
complicated VIII and IX models. For this reason, and for simplicity
we will restrict our attention to the I and II models.

The space-time metric of the Bianchi models can be written as,
\begin{equation}
ds^2=-dt^2+ \sum_{i,j=1}^3 g_{ij} \omega^i \omega^j
\end{equation}
and for simplicity we will only consider diagonal models. The
$\omega$'s are one forms that are invariant under the isometries.
Following Misner \cite{Misner} we introduce the following
coordinatization of the three metric, $g_{ii}=\exp(2\beta^i)$ and we
define $\beta^0,\beta^+,\beta^-$ such that,
\begin{eqnarray}
\beta^1&=&\bar{\beta}^0+\bar{\beta}^++\sqrt{3}\bar{\beta}^-,\\
\beta^2&=&\bar{\beta}^0+\bar{\beta}^+-\sqrt{3}\bar{\beta}^-,\\
\beta^3&=&\bar{\beta}^0-2\bar{\beta}^+.
\end{eqnarray}

We then perform the coordinate transformation \cite{AsUgTa},
\begin{eqnarray}
\beta^0&=& {\sqrt{3} \over 3} (2 \bar{\beta}^0-\bar{\beta}^+)\\
\beta^+ &=& {\sqrt{3} \over 3} (- \bar{\beta}^0+2 \bar{\beta}^+)\\
\beta^- &=& \bar{\beta}^- .
\end{eqnarray}

In terms of these variables one can construct a canonical formulation
for these models with action,
\begin{equation}
S=\int dt \left[p^0\dot{\beta^0} +p^+\dot{\beta^+}+p^-\dot{\beta^-}  
+N \left(-(p^0)^2+(p^+)^2+(p^-)^2+ U(\beta^+,\beta^-,\beta^0)\right)\right].
\end{equation}

For Bianchi I, $U(\beta^+,\beta^-,\beta^0)=0$ and for Bianchi II,
$U(\beta^+,\beta^-,\beta^0)=12 \exp(-4 \sqrt{3}\beta^+)$. The models
therefore generically have a Hamiltonian constraint,
\begin{equation}
-(p^0)^2+(p^+)^2+(p^-)^2+ U(\beta^+,\beta^-,\beta^0)=0.
\end{equation}

In the case of the Bianchi I model, the constraint only depends on the
canonical momenta and not on the coordinates. It is also known that there
exists a canonical transformation that makes constraint only depend on the
momenta in the Bianchi II model. If the constraint depends only on half of
the canonical variables (either momenta or coordinates) the evolution
equations preserve the constraints exactly and the Lagrange multipliers
are not determined. This follows trivially from the fact that if one
has $H(p^i)=0$ the equation of motion following from the canonical
transformation generating evolution include $p^i_{n+1}=p^i_n$ and the
constraint is automatically preserved without fixing the Lagrange
multipliers.
Therefore one obtains a traditional discretization of
the theory, unlike the ones we are discussing in this paper. 

It is interesting to note that in both the Bianchi I and Bianchi II 
models one could choose variables where the constraint is both a function
of coordinates and momenta. In the Bianchi I case the Ashtekar variables
\cite{kodama2}
are an example where this happens. In the Bianchi II case it happens in
the variables we have been discussing. From these variables one can
construct consistent discretizations that determine the Lagrange multipliers.
This exhibits the type of ambiguities that one faces when discretizing 
a continuum model.

We will not present a complete discussion of the Bianchi II model, but 
we would like to highlight some features of interest. The discretized
Lagrangian is,
\begin{equation}
L(n,n+1) =\sum_{a=0,+,-} p^a_n (\beta^a_{n+1}-\beta^a_n)- N_n ((p^0)^2-(p^+)^2-
(p^-)^2)+12 N_n \exp\left(-4\sqrt{3} \beta^+_n\right),
\end{equation}

The resulting equations are,
\begin{eqnarray}
P^0_{n+1} &=&P^0_n\\
P^-_{n+1} &=&P^-_n\\
P^+_{n+1} &=&\epsilon \sqrt{(P^0_n)^2-(P^-_n)^2-12 \exp(-4 \sqrt{3} \beta^+_n)}\equiv
\epsilon\sqrt{\Delta_n}\\
\beta^-_{n+1} &=& \beta^-_n - N_n P^-_{n+1}\\
\beta^0_{n+1} &=& \beta^0_n + N_n P^0_{n+1}\\
\beta^+_{n+1} &=& \beta^+_n - N_n P^+_{n+1}\\
N_n &=& {P^+_n - P^+_{n+1} \over 24 \sqrt{3} \exp(-4 \sqrt{3} \beta^+_n)}
\end{eqnarray}
where as before one first has to define variables canonically conjugate to 
$\beta^+, p^+$, etc. We have written the equations replacing the momenta
canonically conjugate to $\beta$ and relabeled $p^0_n$ as $P^0_{n+1}$.
We have also solved the constraint for the Lagrange multiplier in the 
last equation.

We choose $\epsilon=-1$ and start with initial data at $n=0$ such that
$\Delta_0>0$ and $N_0>0$, $P^0_0>0$. Examining the discrete equations,
one can see that the system starts with expanding $\beta^0$ and
$\beta^+$. $\beta^-$ can either grow or decrease. One can also see
that $\Delta_n>0$ and $N_n>0$ for all $n$.  That is, the solution
expands indefinitely. If one chose $\epsilon=1$ and 
$\Delta_0>0$ and $N_0>0$, $P^0_0<0$, then the universe starts by 
contracting in $\beta^0$, expanding in $\beta^+$. The expansion continues
indefinitely without changes of signs in the variables. Checking the
expression of the three-volume $V_n=\exp(\sqrt{3}[2\beta^0_n+\beta^+_n])$,
one can see that the volume goes to zero at $n\to\infty$. Therefore the
discrete theory has a singularity. Notice that the unavoidability of the
singularity in the discrete theory is related to the fact that in the
continuum the singularity is at the edge of the domain, therefore it
will always be present if one discretizes. One could have chosen a 
set of variables in which the singularity occurs in a point interior to
the domain (for instance, the usual metric variables with appropriate
time parameterization) and then the 
discrete theory could have avoided the singularity as in previous models
we considered. Notice that it is not guaranteed that having the singularity
within the domain will avoid the singularity. Essentially what has to happen
for the singularity to be present in the discrete theory is that it happens
at an accumulation point of grid points, therefore the discrete theory cannot
``walk over'' the singularity.

To show that this is actually possible for this model, let us consider the 
following canonical transformation \cite{AsUgTa},
\begin{eqnarray}
\tilde{P}^+_n &=&\sqrt{(P^+_n)^2+ 12 \exp(-4\sqrt{3} \beta^+_n)},\\
\tilde{\beta}^+_n &=&{\sqrt{3}\over 6} \left\{\left[-P^+_n+\sqrt{(P^+_n)^2+12
\exp(-4\sqrt{3} \beta^+_n)}\right]-\log\left(2 \sqrt{3} \exp(-2\sqrt{3} \beta^+_n)
\right)\right\}.
\end{eqnarray}

The action now reads,
\begin{equation}
S=\int dt \left[p^0\dot{\beta^0} +p^+\dot{\tilde{\beta}^+}+p^-\dot{\beta^-}  
+N \left(-(p^0)^2+(\tilde{p}^+)^2+(p^-)^2\right) \right].
\end{equation}
This action, which has a constraint that is only a function of the momenta,
is an example of the ambiguities present when constructing the discrete theory.
If one discretizes this form of the action, the Lagrange multiplier $N$ is not
determined by the discrete equations and the discrete theory has a completely
different nature than the ones we have considered before. These kinds of 
ambiguities are unavoidable, it is well known that infinitely many discrete
theories can approximate the same continuum theory.

Let us now consider the canonical transformation generated by  $F_2=\exp(\beta^0) p^T$,
such that the new coordinate $T=\exp(\beta^0)$ and $p^0=T P^T$. In terms of
the variables $(T,\beta_\pm)$ and their conjugate momenta $(p^T,p^\pm)$ the 
action now reads,
\begin{equation}
S=\int dt \left[p^T\dot{T} +p^+\dot{\tilde{\beta}^+}+p^-\dot{\beta^-}  
+N \left(-T^2 (p^T)^2+(\tilde{p}^+)^2+(p^-)^2\right) \right].
\end{equation}

The resulting theory, upon discretization exhibits the same behavior as 
the Friedmann model with $m=0$ we discussed before, in particular $T_{n+1}=a T_n$
with $a$ a constant. If $a$ is large enough the universe (run backwards) at some
point departs significantly from the continuum behavior and instead of contracting,
starts to expand, thus avoiding the singularity, as in the case of the Friedmann
model.

Summarizing, we see that the behavior of the Bianchi models is similar to the
behavior of the Friedmann models we discussed in the previous section. It would
be interesting to study the Bianchi models using Ashtekar's new variables. 
Such variables have several attractive features. The singularity happens within
the domain, so it is likely to be avoided in the discrete theory, as it was in 
the case of the Friedmann model with $\Lambda$ and a very massive scalar field.

\section{Conclusions}

We have studied the recently introduced ``consistent discretization''
approach to general relativity in the case of cosmological
models. This helps to clarify, through examples, several questions
that can be raised concerning that approach. First and foremost, the
unusual property of having the Lagrange multipliers determined
operates without difficulty in the models considered.  We never
encounter that the equations imply that the lapse is complex or that
it is multi-valued during evolution. Moreover, the issue of how the
discrete theory achieves a continuum limit is illuminated. We see that
the possibility of having a continuum limit depends on the type of
model and of discretization chosen. Some models exhibit continuum
behavior spontaneously as an attractor in a region of their
evolution. In other models, careful tuning of initial data is needed
to approximate the continuum theory. The Big Bang singularity
can be avoided in a generic fashion in some models, whereas in others
it appears as inevitable, and this can be forecasted from the way
the singularity appears in the continuum theory. We also saw explicitly
how different formulations of the continuum theory can, upon discretization,
exhibit radically different behaviors at the level of the discrete theory.
Finally, the quantization of the discrete theory, in spit of the fact that
the latter has no constraints, is non-trivial. One discovers that 
the discrete theories have remnant symmetries that imply the existence of
constants of the motion that are remnants of the observables (``perennials'')
of the continuum theory. This opens several possibilities for the quantization,
relating to how one approaches the ``problem of time''. One can attempt a
straightforward Schr\"odinger quantization of the theory, but the evolution
one obtains is in terms of a ``time'' that has no physical meaning. One
can attempt to ``gauge fix'' the remnant symmetry and one obtains a theory
with a physically well motivated time that can be quantized a la Schr\"odinger,
but it has the drawback that one has artificially decided to treat a variable
as classical, even in regimes in which it should not be. Finally, one can
attempt a relational introduction of the notion of time that is free of 
the many problems that plagued such attempts in the continuum theory and
therefore provides a natural solution to the problem of time in quantum
cosmology.

It is clear that one can only take the examples presented in this paper,
given their simplicity, as first attempts to understand this approach to
discrete general relativity. Further work will be needed to gain confidence
that the approach can be used in more realistic settings. But already the
exploration of this simple models has been a useful laboratory to exhibit
the kind of behaviors that one may encounter in the discretized theories.

\acknowledgements 

We wish to thank Martin Bojowald, Karel Kucha\v{r} and Rafael Porto
for discussions.  This work was supported by grants NSF-PHY0090091,
funds of the Horace Hearne Jr.  Institute for Theoretical Physics, the
Fulbright Commission in Montevideo and PEDECIBA (Uruguay).


\begin{references}
\bibitem{loll} R.~Loll, Living Rev.\ Rel.\  {\bf 1}, 13 (1998).
\bibitem{us}
C. Di Bartolo, R. Gambini, J. Pullin,
Classical and Quantum Gravity {\bf 19}, 5475 (2002);
R. Gambini, J. Pullin, gr-qc/0206055 to appear in Phys. Rev. Lett.
\bibitem{TDLee} T. D. Lee, 
in ``How far are we from the gauge forces''  
Antonio Zichichi, ed. Plenum Press, (1985)
\bibitem{kodama}
H.~Kodama,
Phys.\ Rev.\ D {\bf 42}, 2548 (1990).
\bibitem{Smolin} L. Smolin ``The life of the cosmos'', Oxford University
Press (1999).
\bibitem{Dirac} P. A. M. Dirac, Z. Phys. Sow. Band 3, Heft 1 (1933),
reprinted in ``Selected Papers on Quantum Electrodynamics'', J. Schwinger,
ed., Dover, New York (1958); see also P. Ramond ``Field Theory:
A Modern Primer'', Benjamin/Cummings, Reading, MA (1981)  p 72.
\bibitem{Kuchar} K. Kucha\v{r}, ``Time and interpretations of quantum gravity'',
in ``Proceedings of the 4th Canadian conference on general relativity and
relativistic astrophysics'', G. Kunstatter, D. Vincent, J. Williams
(editors), World Scientific, Singapore (1992).
\bibitem{Bojobojo}M.~Bojowald Phys.\ Rev.\ Lett.\  {\bf 87}, 121301 (2001).
\bibitem{Dewitt} B.~S.~Dewitt, Phys.\ Rev.\ {\bf 160}, 1113 (1967).
\bibitem{page} D.~N.~Page and W.~K.~Wootters,
Phys.\ Rev.\ D {\bf 27}, 2885 (1983); W. Wooters, Int. J. Theor. Phys.
{\bf 23}, 701 (1984); D. N. Page, ``Clock time and entropy'' in
``Physical origins of time asymmetry'', J. Halliwell, J. Perez-Mercader,
W. Zurek (editors), Cambridge University Press, Cambridge UK, (1992).
\bibitem{porto} R. Gambini, R. Porto, J. Pullin, in preparation.
\bibitem{Misner} C. Misner, in ``Magic without magic'', J. Klauder (editor),
Freeman, San Francisco (1972).
\bibitem{AsUgTa}A.~Ashtekar, R.~Tate and C.~Uggla,
%``Minisuperspaces: Observables And Quantization,''
Int.\ J.\ Mod.\ Phys.\ D {\bf 2}, 15 (1993).
\bibitem{kodama2}H.~Kodama, Prog.\ Theor.\ Phys.\  {\bf 80}, 1024 (1988);
A. Ashtekar and J. Pullin, Ann. Isr. Phys. Soc. {\bf 9}, 65 (1990).
\end{references}
\end{document}